\documentclass{PoS}

\usepackage{wrapfig}
\usepackage{fixltx2e}

\usepackage{tikz}

\title{Theory overview of Heavy Ion collisions}

\ShortTitle{Theory overview of Heavy Ion collisions}

\author{\speaker{Tuomas LAPPI} \\ Department of Physics, 
 P.O. Box 35, 40014 University of Jyv\"askyl\"a, Finland
\\
Helsinki Institute of Physics, P.O. Box 64, 00014 University of Helsinki,
Finland \\
        E-mail: \email{tuomas.v.v.lappi@jyu.fi}}

\abstract{This presentation discusses some recently active topics in the theoretical interpretation of high energy heavy ion collisions at the LHC and at RHIC. We argue that the standard paradigm for understanding the spacetime evolution of the bulk of the matter produced in the collision is provided by viscous relativistic hydrodynamics, which can be used to systematically extract properties of the QCD medium from experimental results. The initial conditions of this hydrodynamical evolution are increasingly well understood in terms of gluon saturation, and can be quantified using Classical Yang-Mills fields and QCD effective kinetic theory. Hard and electromagnetic probes of the plasma provide additional constraints. A particularly fascinating subject are high multiplicity proton-proton and proton-nucleus collisions, where some of the characteristics previously attributed to only nucleus-nucleus collisions have been observed.}

\FullConference{Fourth Annual Large Hadron Collider Physics\\
                 13-18 June 2016\\
                 Lund, Sweden}

    \newcommand{\ud}{\, \mathrm{d}}

\newcommand{\as}{\alpha_{\mathrm{s}}}
\newcommand{\lqcd}{\Lambda_{\mathrm{QCD}}}

\newcommand{\ptt}{{p_T}}

\newcommand{\xt}{{\mathbf{x}_T}}

\newcommand{\qs}{Q_{\textnormal{s}}}

\usepackage{amssymb}
\usepackage{amsmath}

\begin{document}

\section{Introduction}

The topic of this talk is understanding LHC experiments in terms of what has been referred to as the ``Condensed Matter Physics of QCD''. The Lagrangian for Quantum Chromodynamics, the theory of the strong interaction, can be simply written down on one line:
\begin{equation}
 \mathcal{L} = -\frac{1}{4}F^{\mu\nu}_a F^a_{\mu \nu} 
+ \overline{\psi}_i(i D\!\!\!\!\!/ -m_i)\psi.
\end{equation}
In spite of this brevity, the theory leads to a wealth of interesting phenomena. Heavy ion collisions at the LHC and at RHIC create macroscopic (compared to the relevant microscopic length scale given by the inverse temperature) amounts of QCD matter. Studying the properties of this ``QCD condensed matter'' through an active interaction between theory and experiments allows us to gain unique insights to the behavior of the strong interaction.

\begin{figure}[t]
\centerline{
\resizebox{8cm}{!}{
\includegraphics[width=6cm]{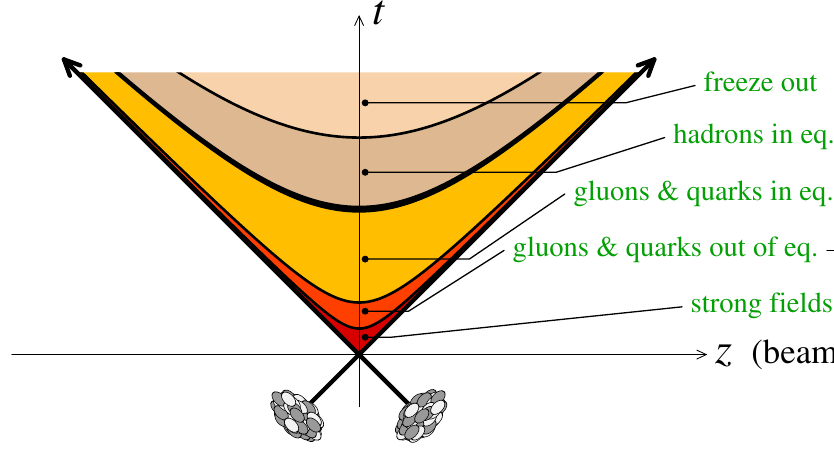} 
\begin{tikzpicture}[overlay]
\filldraw[white] (-5.4,0.7) ellipse (0.6 and 0.2);
\filldraw[white] (-0.4,0.7) ellipse (0.45 and 0.2);
\filldraw[white] (-2.2,0.8) rectangle (0,1.15);
\node[anchor=west] at(-2.2,1) {glasma fields};
\filldraw[white] (-2.4,1.25) rectangle (0,1.55);
\node[anchor=west] at(-2.4,1.4) {nonequilibrium partons};
\filldraw[white] (-1.99,1.65) rectangle (-0,2.0);
\node[anchor=west] at(-1.99,1.85) {plasma};
\filldraw[white] (-1.35,2.1) rectangle (0,2.5);
\node[anchor=west] at(-1.35,2.3) {hadron gas};
\filldraw[white] (-1.05,2.6) rectangle (0,2.9);
\node[anchor=west] at(-1.05,2.75) {freezeout};
\end{tikzpicture}
}
}
\caption{Phases of a heavy collision}
\label{fig:spacet}
\end{figure}
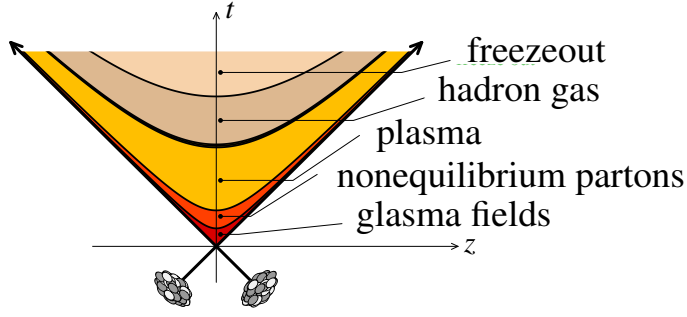
One of the most important properties of QCD is asymptotic freedom: at short enough distances $\ell \lesssim 1/\lqcd$ the interactions between colored particles become weaker. The typical separation between particles in a relativistic thermal system is given by the inverse of the temperature $\ell \sim 1/T$, and one therefore expects colored elementary constituents of hadrons to become weakly coupled at high enough temperatures $T\gtrsim \lqcd$~\cite{Shuryak:1980tp}. Confined bound states, hadrons, can then be expected to melt into a plasma of more elementary colored constituents, quarks and gluons. Already the exponentially increasing spectrum of hadronic resonances points to the existence of a limiting temperature~\cite{Hagedorn:1965st} beyond which hadronic matter is inconsistent. By now QCD is well established as the correct theory of the strong interaction. Its thermodynamics can be solved on the lattice to increasingly high accuracy~\cite{Ratti:2016jgx}. These lattice studies confirm the transition to a deconfined phase (a rapid crossover, not a genuine phase transition in the absence of a  baryochemical potential) at a temperature around 150MeV.

The aim of the heavy ion collision program at LHC and RHIC is to study the properties of this deconfined matter in the laboratory with heavy ion collisions at varying energies. This is an integral part of understanding the different aspects of the fundamental interactions of elementary particles in the standard model. One needs to stress here that the goal is to study the properties of \emph{matter}, and not individual particles. This means that a good theoretical understanding and interpretation of the experimental results requires studying the global characteristics of the bulk of particles produced in a the collision. Thus one is not looking for rare and exceptional events, but the aim is to understand the typical collision process in terms of the properties of the matter produced in it.

The system produced in a heavy ion collision evolves in time and space in multiple phases (see Fig.~\ref{fig:spacet}).  Relating experimental observations to fundamental properties of QCD requires one to build a consistent picture of all of them. Firstly one needs to understand the physics of the relevant degrees of freedom inside the colliding nuclei. One then must calculate how these degrees of freedom can equilibrate to a system close enough to thermal equilibrium that it can be understood in terms of thermodynamic concepts like the ``equation of state''.  The plasma is not static; it lives for a finite amount of time and during this time undergoes a rapid expansion in space into the vacuum surrounding the collision. The standard tool for understanding this spacetime evolution is relativistic hydrodynamics. Finally the plasma cools down enough to transition again into the hadronic phase and flow as particles to the detectors. From all the stages of this process, distinct rare probes (hard partons and electromagnetically interacting particles) can be emitted. They interact with the medium as they pass through it, and give us additional information on the QCD matter that they qent through. This talk will present a very limited and  biased selection of recent advances and open questions in these  different aspects of the theoretical description of a heavy ion collision.

\section{Initial stages and thermalization}

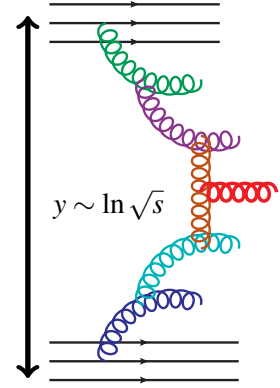
\begin{wrapfigure}{r}{4cm}
\centerline{
\includegraphics[height=5.2cm]{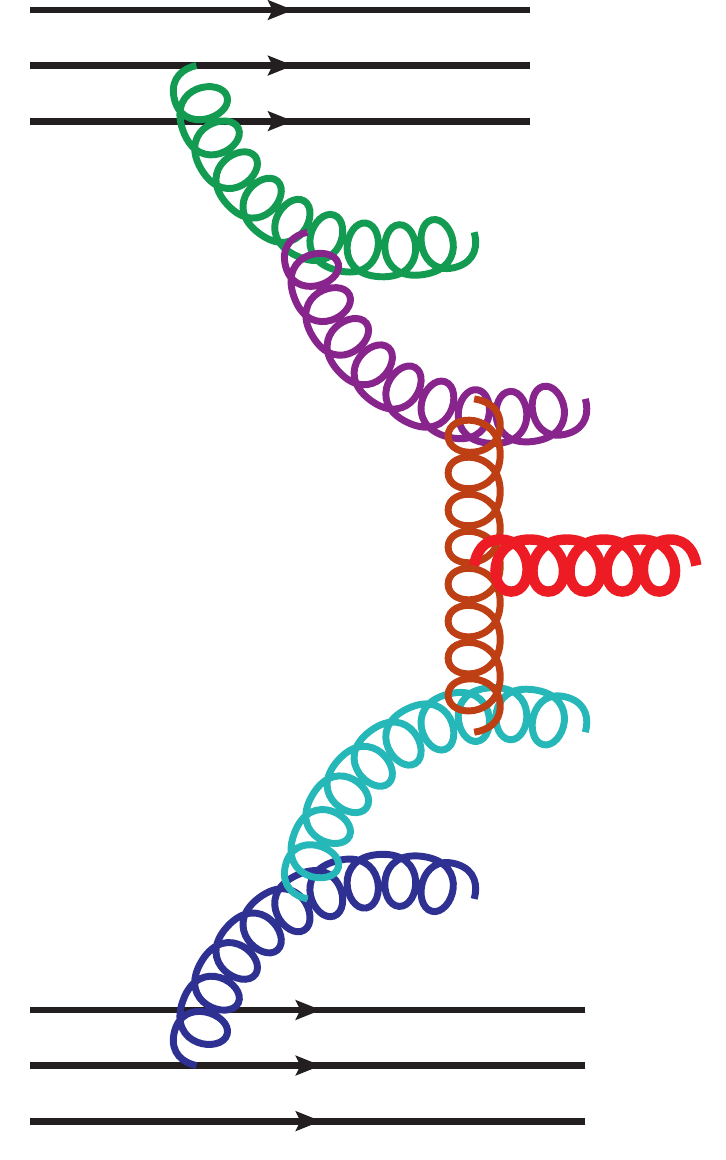}
\begin{tikzpicture}[overlay]
\draw[line width=2pt,<->](-3.5cm,0.2cm) -- (-3.5cm,5cm);
\node[anchor=west]at (-3.3,2.5) {$y\sim \ln \sqrt{s}$};
\end{tikzpicture}
}
\caption{Gluon cascade from the colliding nuclei}
\label{fig:casc}
\end{wrapfigure}
The bulk of the matter forming a quark-gluon plasma in the central rapidity region initially consists of gluons. These gluons are the result of multiple gluon emissions in an exponentially growing cascade that starts from the valence quarks (see Fig.~\ref{fig:casc}). The probability to emit a gluon with a (small) momentum fraction $x$ of its parent quark or gluon is parametrically $\as \ud x/x = \as \ud y$, where $y$ is the rapidity of the gluon. This cascade picture of weak coupling QCD therefore naturally leads to particles being produced in  a boost-invariant way to leading order: at rapidity scales 
$\Delta y\lesssim 1/\as$ the distribution is independent of rapidity. This plateau, also seen experimentally, should be contrasted with a strong coupling picture that generically leads to a stopping of the colliding nuclei followed by an explosion that leads to a more isotropic distribution in the scattering angle.

When the collision energy is large enough, the available phase space for gluon emissions compensates for the smallness of the coupling constant $\as$. Eventually, at a high enough $\sqrt{s}$, the occupation numbers of gluonic states become classical (in the sense of classical fields, not particles), i.e. of order $f(k) \sim A_\mu A_\mu \sim 1/\as$. This leads, even when the coupling is small, to a breakdown of the perturbative expansion in terms of powers of $gA_\mu$. For momenta of the order of some characteristic momentum scale $\qs$ the system is dominated by the nonlinearities; the interaction terms in the Yang-Mills Lagrangian. This typical transverse momentum scale $\qs$ is known as the \emph{saturation scale}; below it the nonlinear interactions of gluons limit the exponential cascade of gluon emissions. At higher energies even more gluons can be emitted; the saturation scale thus grows with energy. At some point $\qs \gg \lqcd$, and a  weak coupling expansion can be used.  Due to the high densities this cannot, however, be organized as a conventional perturbative expansion, but must be thought of as an expansion around a nonperturbatively large classical field.

\begin{wrapfigure}{r}{0.6\textwidth}
 \includegraphics[width=0.6\textwidth]{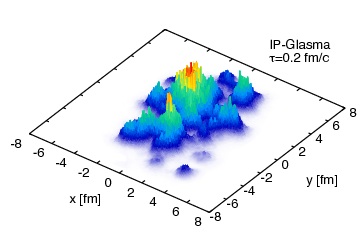}
\caption{Energy density in the transverse plane at $0.2$fm after a heavy ion collision from the IPglasma~\cite{Schenke:2012hg} model, which is based on the color field radiated from a color charge distribution fit to HERA DIS data~\cite{Kowalski:2006hc}.}
\label{fig:ipglasma}
\end{wrapfigure}

Many recent attempts to understand initial particle production in a heavy ion collision are based on the idea of gluon saturation~\cite{Blaizot:1987nc}, either using a classical field calculation or a more perturbative picture. As an example, Fig.~\ref{fig:ipglasma} shows the energy density at a time of $0.2$fm after the collision in the IPglasma model~\cite{Schenke:2012hg}. This is a numerical solution of the classical Yang-Mills equations for the gluon fields radiated from classical color sources, taken from a McLerran-Venugopalan (MV)~\cite{McLerran:1994ni} model Gaussian probability distribution. The magnitude and transverse coordinate dependence of the color charge density is taken from a fit to HERA data~\cite{Kowalski:2006hc}, yielding a prediction without, at least in principle, any free parameters. An interesting feature that sets the IPglasma model apart from many others is the presence of energy density fluctuations also at subnucleon size scales, resulting ultimately from quantum fluctuations in the degrees of freedom in the colliding nuclei.

\begin{wrapfigure}{r}{0.6\textwidth}
\centerline{\includegraphics[width=0.6\textwidth,clip=true,bb = 0 0 257 215]{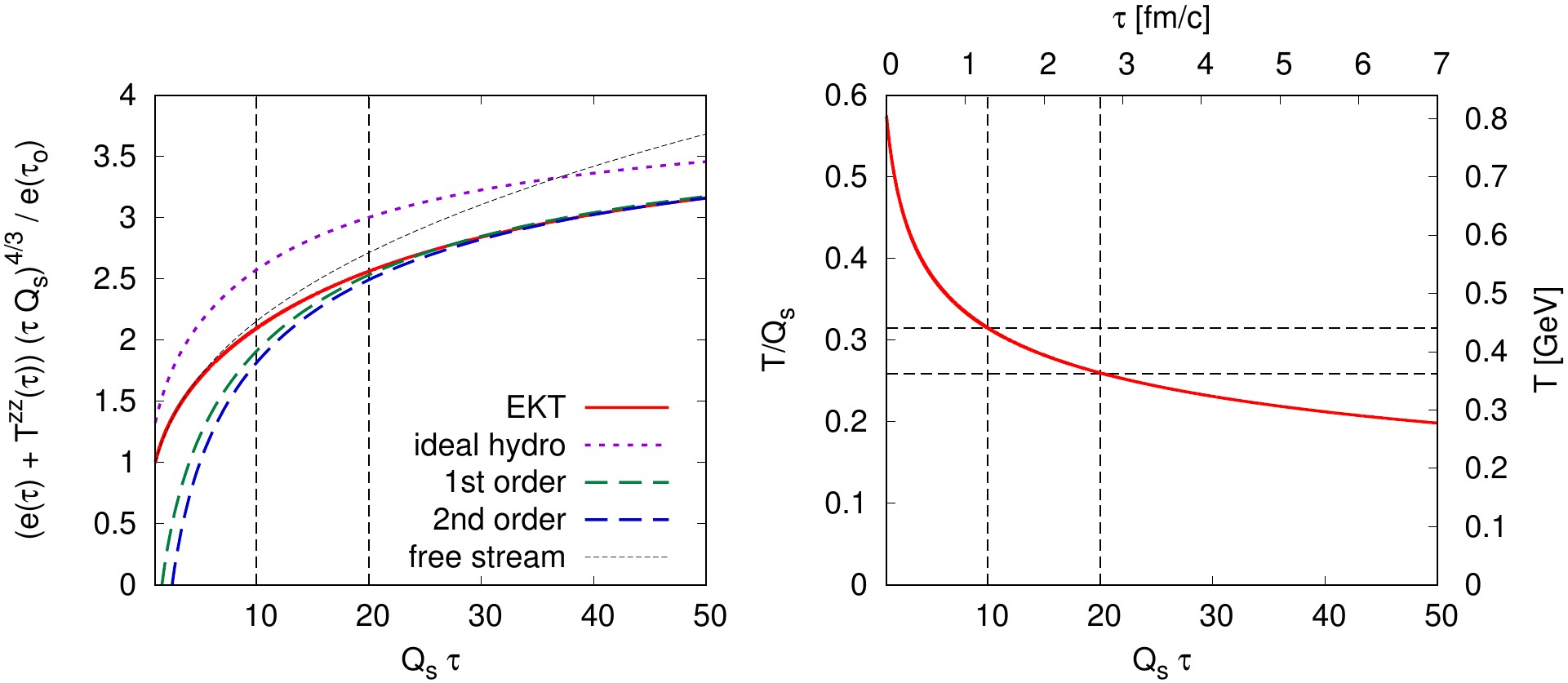}}
\caption{Apropriately scaled combination $T^{00}+T^{zz}$ of diagonal components of the energy momentum tensor: from free streaming particles, viscous hydrodynamics and the QCD effective kinetic theory that interpolates between them, from Ref.~\cite{Keegan:2015avk}. }
\label{fig:interp}
 \end{wrapfigure}
The ``glasma''\cite{Lappi:2006fp} fields in the initial stages of the collisions are, to leading order in the coupling $\as$, boost invariant. This invariance, at the level of fields configurations, corresponds in a quasiparticle picture to a system of gluons with momenta only in the transverse direction, i.e. a very anisotropic system. The expansion of the system in the longitudinal direction tends, in the absence of strong enough interactions, to make this anisotropy even greater due to a redshift of the longitudinal components of the gluon momenta. A great deal of progress has been made quite recently in understanding how this initial state can manage to isotropize into something close enough to local thermal equilirium to be understood in terms of viscous relativistic hydrodynamics. The theoretical framework that allows one to follow this development in the weak coupling picture is provided by the so called QCD effective kinetic theory (EKT)~\cite{Arnold:2002zm}. The EKT includes, in addition to hard $2\to2$ scatterings, also an effective $1 \to 2$ splitting that allows one to resum $n\to n+1$ scatterings that contribute at  the same order in the coupling due to an enhancement by a large collinear logarithm. Figure~\ref{fig:interp} shows the result of a recent EKT calculation~\cite{Kurkela:2015qoa,Keegan:2015avk} in a longitudinally expanding system that starts from the very anisotropic glasma initial state and matches smoothly to a viscous hydrodynamical description at a relatively early time. This calculation has now finally enabled us to follow, in a first principles QCD calculation, the time evolution of the system from the initial classical color fields to plasma close enough to local thermal equilibrium to be describable in terms of viscous hydrodynamics.

\begin{figure}[t]
\centerline{\includegraphics[height=6cm]{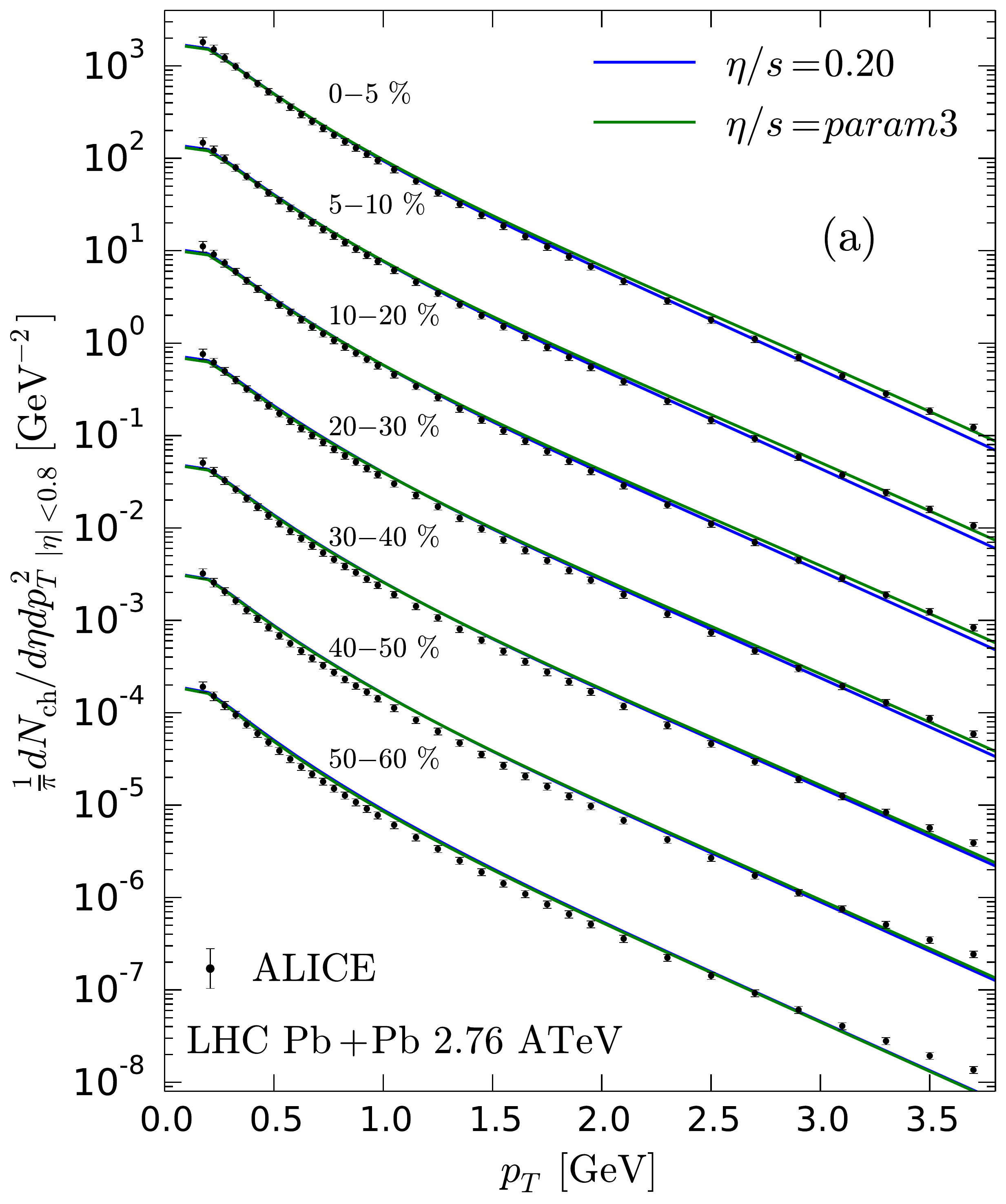}
\includegraphics[height=6cm]{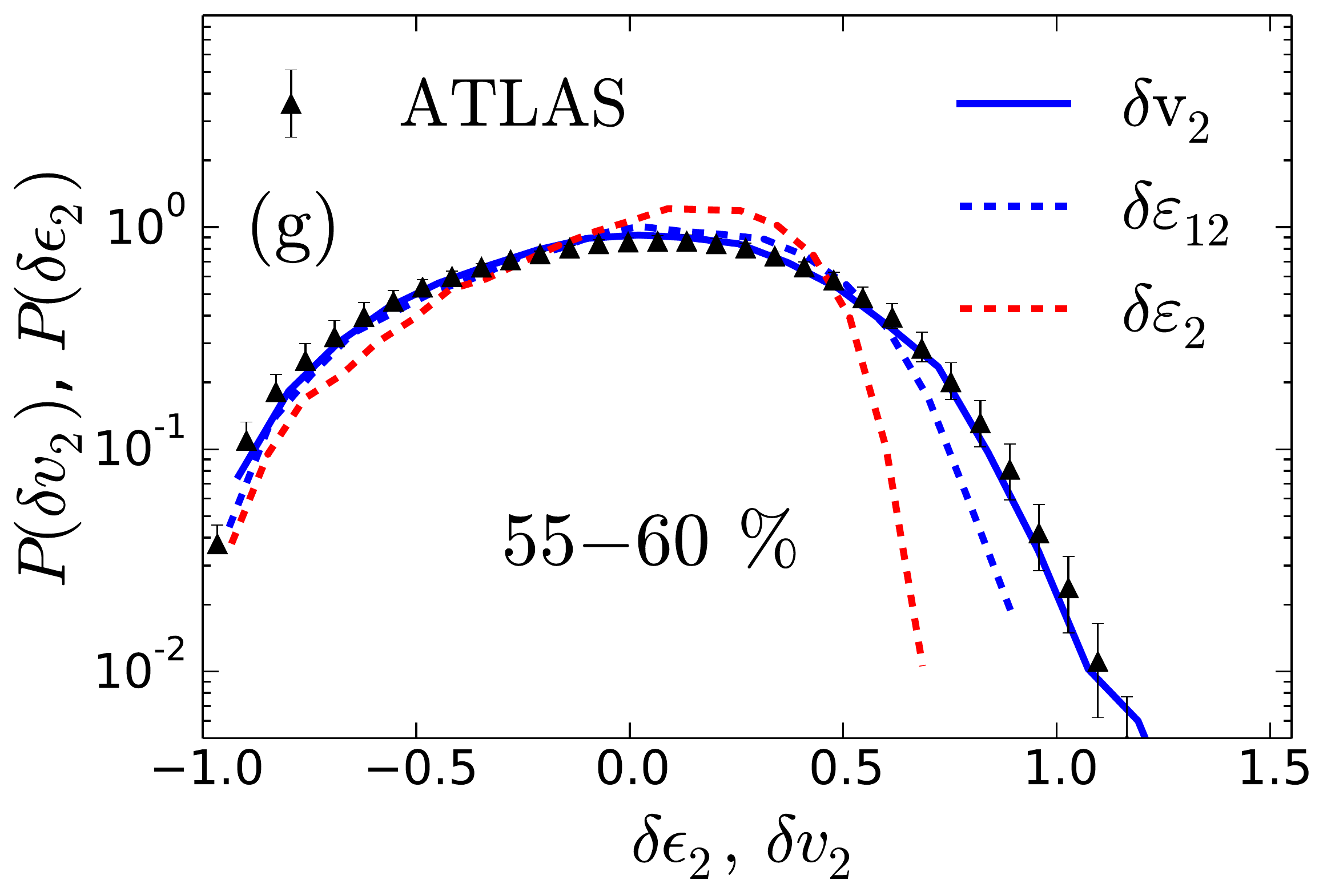}}
\centerline{
\includegraphics[height=6cm]{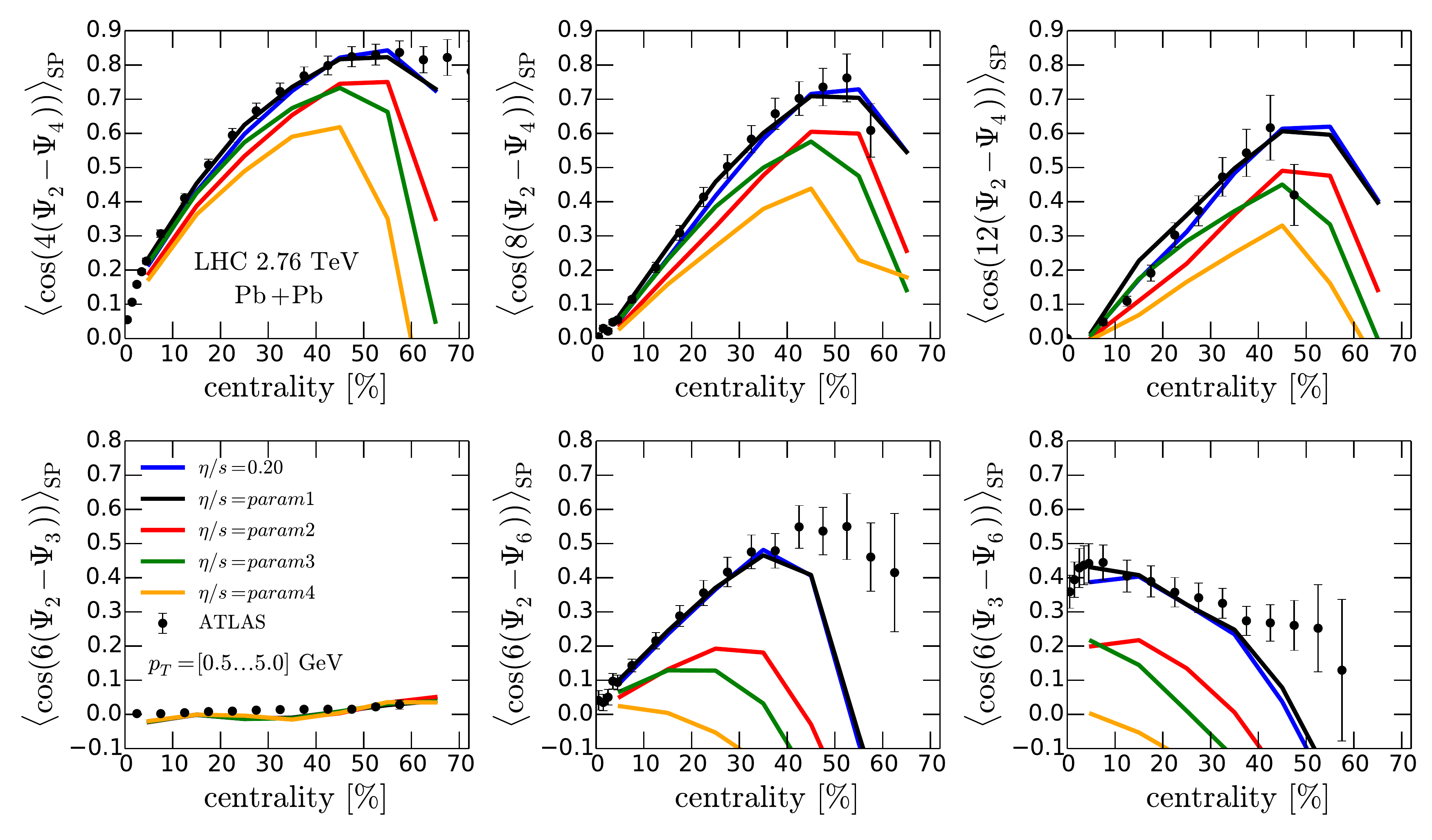}
}
\caption{Comparisons of ALICE data with a hydrodynamical model~\cite{Niemi:2015qia} for specific parametrizations of the temperature-dependent viscosity of QCD matter $\eta/s$, top left: charged particle $\ptt$-spectra for different collision centralities, top right: probability distributions for azimuthal harmonic coefficients of the momentum distributions ``$v_n$'s'' and bottom: correlations between the directions of different ``reaction planes,'' i.e. the directions in the transverse plane determined by different azimutghal Fourier harmonics of particle correlations.}
\label{fig:ekrthydrocomp}
\end{figure}

\section{Global characterization of heavy ion event with hydrodynamics}
\label{sec:hydro}

Hydrodynamical simulations are the most common and successful way of obtaining a global theoretical description of heavy ion events. In addition to the initial conditions described above, what is needed are two additional physical ingredients. Firstly, one solves the equations of motion of hydrodynamics, which  consist simply of the conservation of energy and momentum:
\begin{equation}
\partial_\mu T^{\mu \nu}=0 . 
\end{equation}
Secondly, to close this system of equations, one  needs constituent relations between the components of the energy momentum tensor. For ideal hydrodynamics, i.e. a fluid in full local thermal equilibrium, the only one required is the equation of state $p(\varepsilon,\mu)$, expressing the pressure as a function of energy density and conserved charges (most importantly the baryon number). In viscous hydrodynamics the fluid can deviate from a local thermal equilibrium, and one also needs transport coefficients that relate these deviations to gradients of the fluid velocity. These are all in calculable (although not easily) from QCD, and thus fundamental properties of the standard model. The system is then evolved in time until it expands and cools enough to freeze out into noninteracting particles.

The freezeout from a hydrodynamical simulation  generates particle distributions that can directly be compared to experimental ones. Thus the predictive power of the framework is very strong, as it allows for a simultaneous comparison of theory to experiment for a multitude of soft observables. A modern hydrodynamical description (Fig~\ref{fig:ekrthydrocomp} shows the results of Ref.~\cite{Niemi:2015qia} as an example) should, for example, describe the $\ptt$-spectra (both the yields and mean transverse momenta) of hadrons, the azimuthal $\cos n \varphi$ Fourier harmonics of multiparticle correlations  ``$v_n$'s'', and correlations between azimuthal angles corresponding to the planes of harmonics with different harmonic number $n$. All of this information can be calculated  and measured for identified hadrons, and for different centrality classes of collisions. All in all, the quantity of experimental observables that can be described simultaneously is quite remarkable.

\begin{wrapfigure}{R}{0.55\textwidth}
 \includegraphics[width=0.55\textwidth]{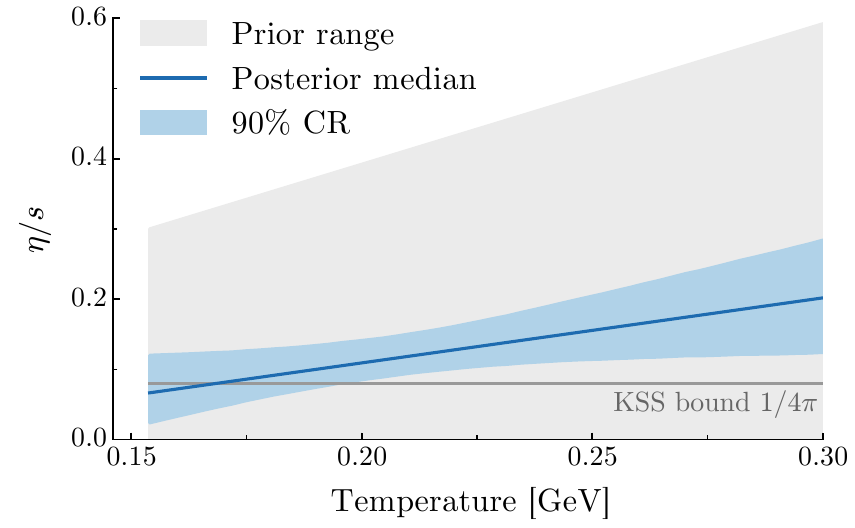}
\caption{Estimate for the QCD matter viscosity from \cite{Bernhard:2016tnd}.}
\label{fig:morelandvisc}
\end{wrapfigure}

A current trend in hydrodynamical simulations is to move from demonstrating that one can describe the experimental data towards extracting QCD medium properties from it in a statistically well defined unbiased way. A recent example of this probram is the Bayesian global fit strategy demonstrated in Ref.~\cite{Bernhard:2016tnd}. A particular feature this work is the rather flexible model-independent parametrization of the initial entropy density in the transverse plane, which is taken as a generalized mean
\begin{equation}
s(\xt) \sim \left(\frac{\left(T_A(\xt)\right)^p + \left(T_B(\xt)\right)^p}{2}\right)^{\frac{1}{p}}
\end{equation}
of the densities of the colliding nuclei $T_{A,B}(\xt)$, which are fluctuating superpositions of individual nuclei within the nucleus. The fit to experimental heavy ion data favor values around $p\approx 0$, which is consistent with semihard initial particle production as in the gluon saturation picture. The current estimate for the viscosity is shown in Fig.~\ref{fig:morelandvisc}. At this stage it should be regarded as a proof-of-principle result that the statistical framework can indeed converge to a meaningful result. The inclusion of more experimental data will allow for a more flexible fit parametrization for the temperature dependent viscosity, and we are likely to see further improvements from these calculations. 

\begin{figure}[t]
\resizebox{\textwidth}{!}{
\includegraphics[height=5cm]{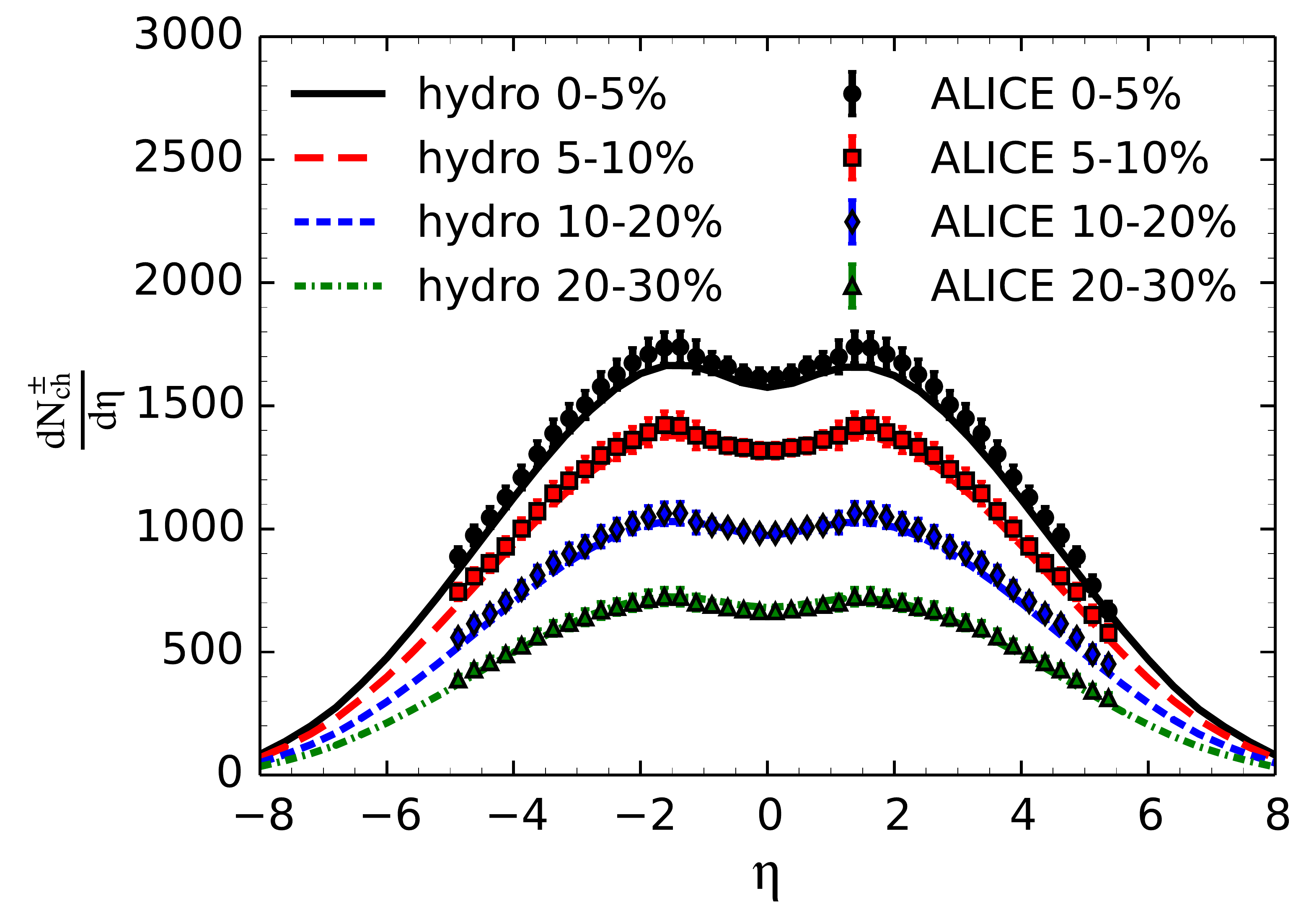}
\includegraphics[height=5cm]{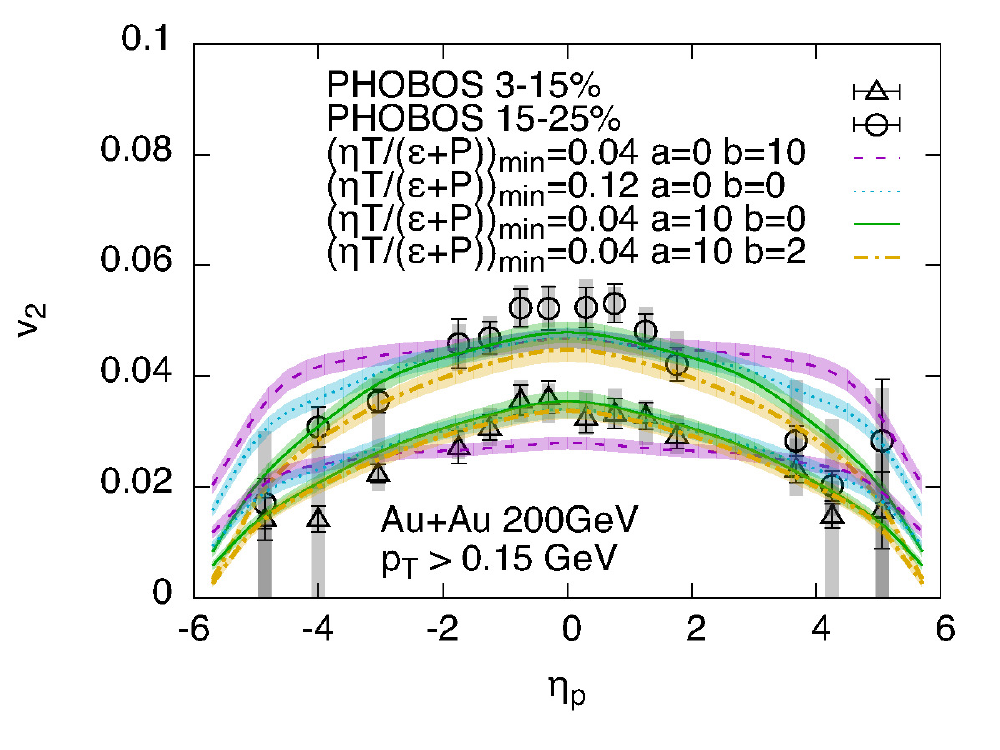}
}
\caption{Hydrodynamical calculations of the charged multiplicity~\cite{Pang:2015zrq} compared to ALICE data and of the elliptic flow coefficient $v_2$~\cite{Denicol:2015nhu} compaerd to PHOBOS as a function of rapidity.}
\label{fig:rap}
\end{figure}

\begin{wrapfigure}{R}{0.5\textwidth}
\includegraphics[width=0.5\textwidth]{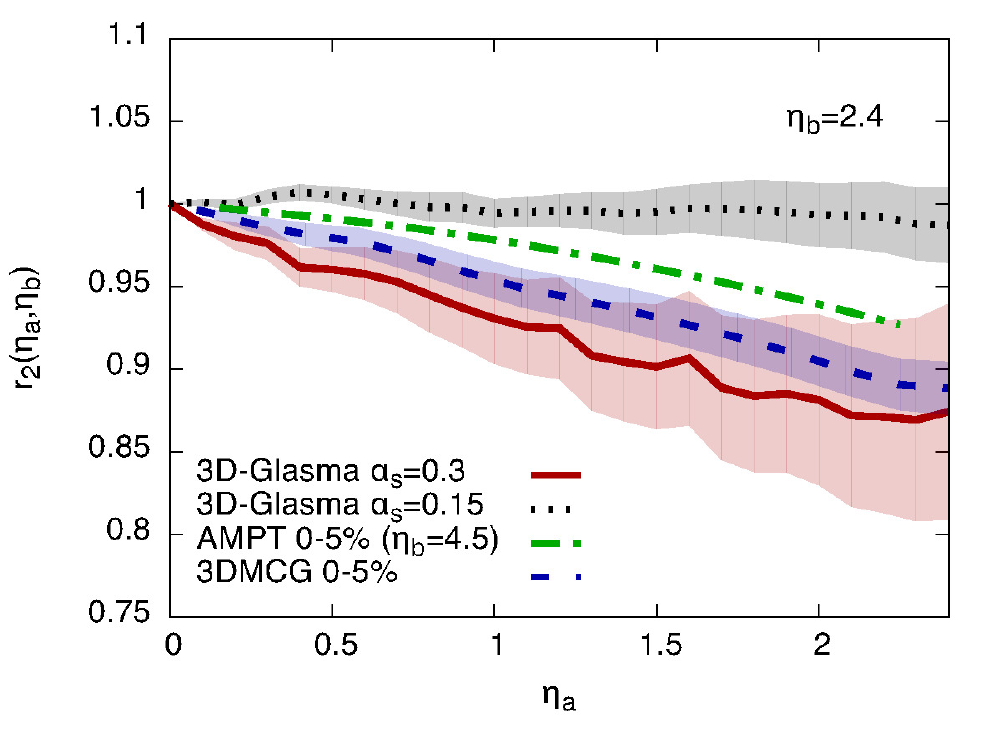}
\caption{
Decorrelation between event planes (i.e. the directions of azimuthal anisotropy in the transverse plane) as a function of rapidity separation, from ~\cite{Schenke:2016ksl}.
}
\label{fig:rapcorr}
 \end{wrapfigure}

In spite of encouraging successes, the hydrodynamical description of heavy ion collisions still has room for improvement on the theoretical front, in order to exploit an even wider set of experimental data. One yet poorly understood aspect of the initial conditions is the dependence on rapidity, and in particular correlations between different rapidities. Calculations for the rapidity dependence are mostly only available in the form of parametric estimates or string-type models of the initial conditions (e.g. \cite{Bozek:2015swa,Pang:2015zrq,Denicol:2015nhu}). A few recent predictions from these on the rapidity dependence of the particle multiplicity and second azimuthal anisotropy coefficent $v_2$ are shown in Fig.~\ref{fig:rap}. Currently we have some idea of how to calculate correlations at large rapidity separations from a microscopic theory~\cite{Iancu:2013uva}, but no few concrete realizations of these ideas (see however~\cite{Gelis:2008sz,Lappi:2012gg,Schenke:2016ksl} and Fig~\ref{fig:rapcorr}).

Including viscosity in the hydrodynamical calculation enables one to exted it further from local thermal equilibrium. Hydrodynamics only cares about the energy momentum tensor and is agnostic concerning the internal degrees of freedom that make it up. But at the freezeout hypersurface the  energy-momentum tensor has to be converted into a distribution of hadrons. In thermal equilibrium there is a unique way to do this, but the deviations from equilibrium make the situation more complicated. The question of these ``viscous corrections to freezeout'' in particular limits our ability to extend the hydrodynamical description to higher momentum particles, where these corrections are large. At some point in the high momentum tail of the distribution, hadrons cease to be a part of the medium and must be understood as nonthermal hard probes interacting with it.

\section{Small systems and the limits of collectivity}

The success of the hydrodynamical description in heavy ion collisions has led the community to look for the limits of the collective behavior.  At the higher collision energies, and with the high luminosity allowing one to select extremely high multiplicity events in proton-proton and proton-nucleus collisions, the LHC has produced some of its most fascinating recent results. These systems had previously been though of as ``small'' ones, where collective effects and quark gluon plasma formation would not be expected. There is a lively discussion going on in the field about the limits of collectivity: how small of a system can one still meaningfully describe with hydrodynamics? Are all the signals in larger collision systems  that have heretofore been attributed to collectivitity indeed such, if they also appear in ever smaller systems? In this conference we have also heard in several talks,  that another community is now approaching the same questions from a different point of view: Monte Carlo descriptions that have so far worked well for soft particle production in  proton-proton collisions are increasingly having problems describing the very high multiplicity events.

\begin{figure}[t]
\includegraphics[width=\textwidth]{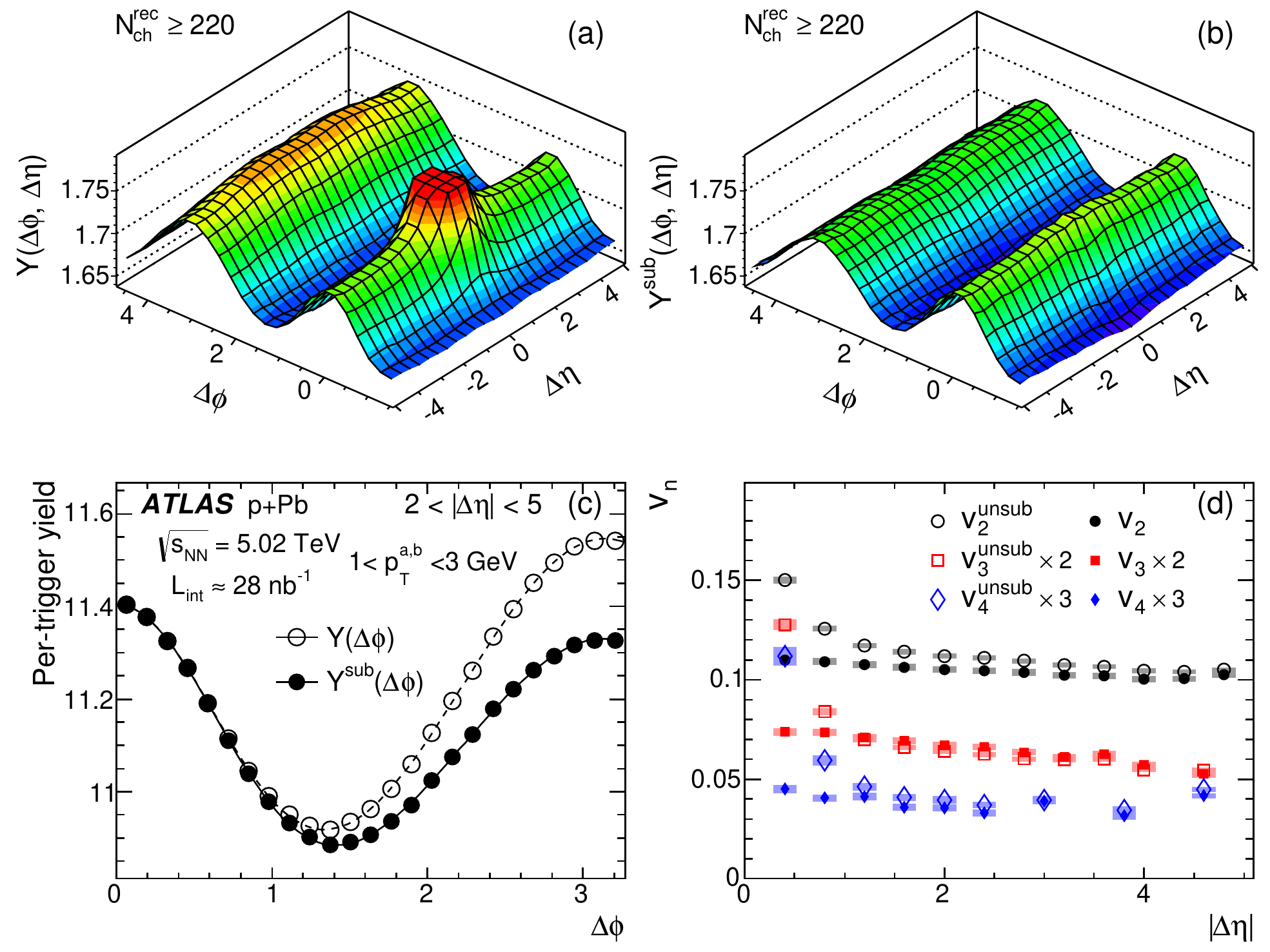}
\caption{Two particle correlation analyzed either as a two-dimensional correlation function (yield per trigger) in the azimuthal angle -- pseudorapidity plane (top, bottom left) or as  flow coefficients $v_n$ (i.e. the coefficient of a $\cos (n\varphi)$ asymmetry in a single particle distribution, bottom right), from ATLAS~\cite{Aad:2014lta}.
}
\label{fig:atlaspaflow} 
\end{figure}

\begin{figure}[t]
\centerline{\includegraphics[width=0.6\textwidth]{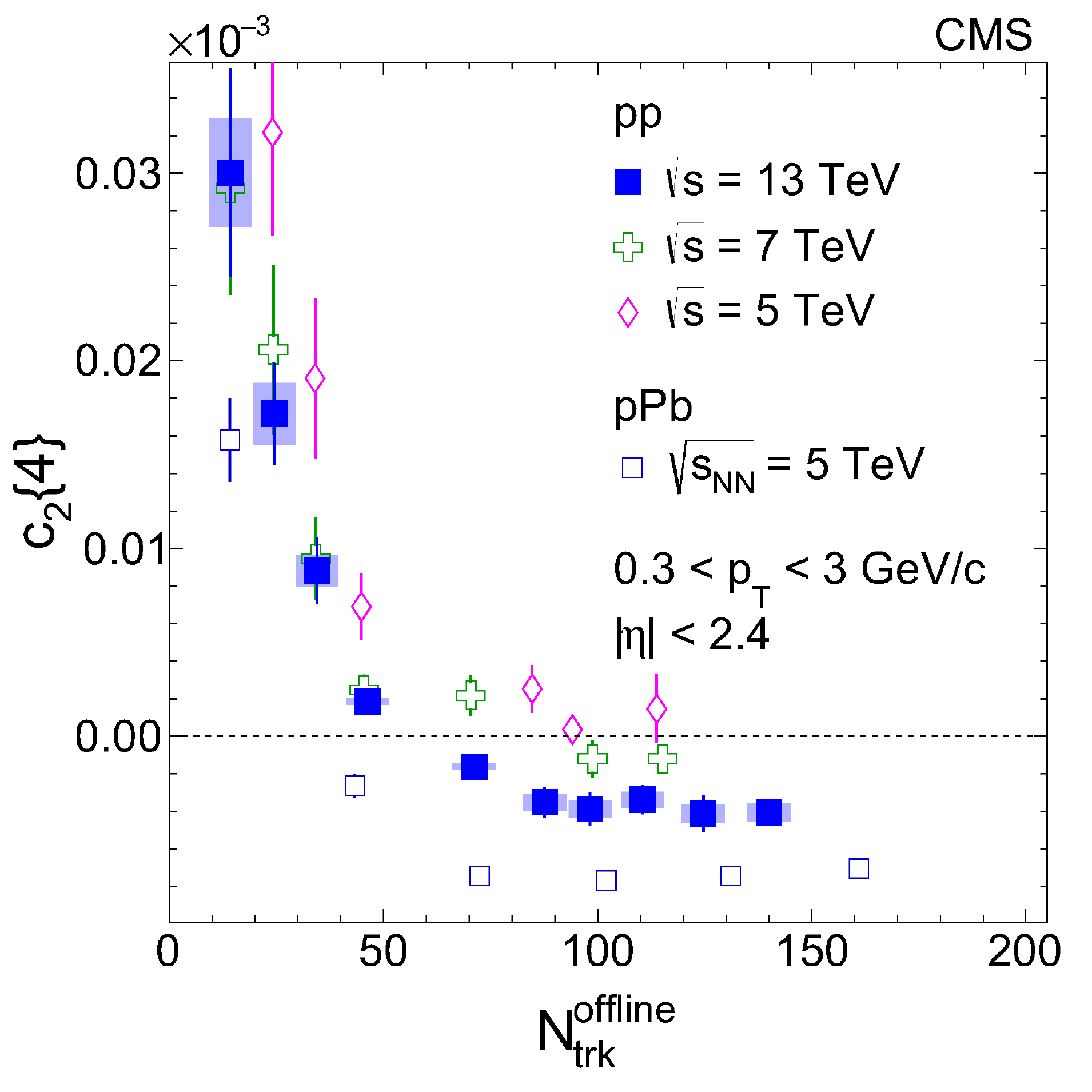}}
\caption{A recent measurement of the four-particle cumulant $c_2\{4\}$, where the subscript refers to the harmonic $\cos (2\varphi)$ and the coefficient is extracted from a four-particle correlation function to suppress the effect of dijet-like back-to-backl correlations. Note the sign change as a function of total charged multiplicity in the event, denoting a change from a two-particle-correlation dominated regime with $c_2\{4\} >0$ to a multiparticle correlation involving most of the particles produced in the collision  $c_2\{4\} <0$. Measured by the CMS collaboration ~\cite{Khachatryan:2016txc}.
}\label{fig:cmsppc24}
\end{figure}

Figures \ref{fig:atlaspaflow} and \ref{fig:cmsppc24} show some recent experimental results from small collision systems, in both cases concentrating on ``elliptic flow'' coefficients in high multiplicity proton-nucleus and proton-proton collisions. These correlations are usually thought of as hallmark signals of collective final state interactions, i.e. the presence of a genuine matter whose spatial size and lifetime are large compared to the microscopic scales at this energy. These experimental observations have prompted questions on alternative mechanisms  in terms of multiparticle correlations present already in the wavefunctions of the colliding particles. Indeed it has become clear that there are strong angular correlations between the color fields (gluons) already  in the incoming nuclei~\cite{Dumitru:2015gaa}. These correlations then become visible in the particles that are produced in the collision~\cite{Dumitru:2010iy,Kovner:2010xk}, and could produce an effect that mimics a hydrodynamical flow correlation.

In the standard hydrodynamical picture elliptic flow, or more generally multiparticle azimuthal correlations, are generated in the following way. The initial energy density of the plasma is not radially symmetric. For example, an in an off-central collision (i.e. with a nonzero impact parameter between the centers of the collinding nuclei)  the overlap in the transverse plane of the two nuclei has an almond-like elliptical shape. This almond-like shape is parametrized in terms of the initial state ellipticity $\varepsilon_2$, which measures the $\cos 2 \varphi$-deformation of the distribution of matter in \emph{space}. More generally, fluctuations of the positions of the individual nucleons within a nucleus (including these fluctuations has been one of the most important advances in hydrodynamical calculations in recent years) lead to fluctuations in the energy density of the plasma in the transverse plane. The fluctuations can be parametrized in terms of a power spectrum of harmonic coefficients $\varepsilon_n$. The presence of energy density fluctuations means that there are temperature and pressure gradients in the transverse plane.  When this energy density consists of collectively interacting matter (i.e. behaves hydrodynamically) the anisotropic pressure gradients lead to anisotropic forces acting on the matter, which in turn leads to an anosotropic acceleration. In the initial state, the momentum distributions of produced particles must be locally isotropic, at least in the case of known particle production mechanisms in QCD. During the time evolution of the plasma, however, the anisotropic acceleration of the fluid cells of  leads to an anisotropy in the momentum distributions, and eventually in the detected particles. This anisotropy is parametrized by the variable $v_n$, the coefficient of the $\cos n \varphi$ term in an azimuthal angle Fourier series of the momentum distribution of all the particles in one event.

Thus in the hydrodynamical picture one starts from an anisotropy in coordinate space which, due to a macroscopic lifetime of the system spent under the effect of collective interactions, turns into a momentum-space anisotropy involving \emph{all} the particles in the event. This is to be contrasted with correlations from momentum conservation in e.g. dijet events. Here particles are also correlated with a  $\cos 2 \varphi$ pattern, but the correlation only involves particles originating from the same dijet. In a high-multiplicity collision this latter kind of an effect is present in two-particle correlations, but is completely washed out by the combinatorics when all the produced particles  are correlated with each other e.g. in a 4-particle cumulant methdod (thus the importance of multiparticle cumulants, see Fig.~\ref{fig:cmsppc24}). 

In a collision of two lead ions the size of the plasma fireball is large, of the order of the nuclear radius which is much larger than the inverse temperature (or mean free path) in the plasma. 
For large collision systems the the initial geometry  as parametrized by $\varepsilon_n$'s is controlled by the well known density of nucleons inside a nucleus and relatively independent of the microscopic dynamics. 
Thus we can with a good confidence attribute the success of hydrodynamical calculations to the presence of genuine collectivity in the system. The similar effects seen in collisions of smaller systems have led to several hydrodynamical calculations to explain them in a similar framework~\cite{Bozek:2011if,Werner:2013ipa,Bozek:2014wpa}. Although the ``hydrodynamical response function'' transforming the initial coordinate space asymmetries $\varepsilon_n$ to  final state momentum space ones is by now well known also here, the initial coordinate space asymmetry itself is not understood well enough. 
In contrast to large systems, for small systems the geometry depends on length scales smaller then the nucleon size, and thus on the the microscopical particle production mechanism. For small systems this generates differences of more than a factor of 2 in the coefficients $\varepsilon_n$ between different commonly used models of initial particle production~\cite{Bzdak:2013zma}.

Fortunately there are ways to solve this puzzle, with additional efforts from theory to exploit a broader array of experimental data. One important aspect of this is to use higher order cumulants (e.g. the 4-particle correlation results such as shown in Fig.~\ref{fig:cmsppc24}) to descriminate flow- and nonflow effects. Any non-hydrodynamical mechanism purporting to explain the flow-like signatures in terms of initial state gluon correlations in momentum space must also be able to reproduce the structure of higher cumulants, which could provide a very strong discriminating factor for these calculations.
For a more detailed discussion of different implementations~\cite{Dumitru:2014dra,Dumitru:2014vka,Lappi:2015vha,Schenke:2015aqa} of initial state angular correlations see e.g Ref.~\cite{Lappi:2015vta}.
Another experimental result that is only now starting to be utilized are correlations between different rapidities. In a purely geometrical initial condition model for hydrodynamics, particle production at all rapidities is controlled by the same positions of the original sources of particles, leading to an infinite range correlation in the rapidity direction. As discussed in Sec.~\ref{sec:hydro}, more microscopical models will lead to a decorrelation between different rapidities and can be used to distinguish between mechanisms generating  azimuthal correlations in coordinate or momentum space. Finally, the statistical hadronization from near thermal equilibrium, as in a hydrodynamical picture, produces a specific ``mass ordering'' between flow signals (and mean transverse momenta etc) for identified hadrons. This is basically due to the fact that for a given flow velocity (common to all particle species) heavier particles have a larger momentum. This mass ordering is seen in the $v_n$ data for small systems, and has been taken as a clear sign of collective behavior. The situation is less clear if the system is far from equilibrium, because describing hadronization is more difficult. A recent preliminary study~\cite{Schenke:2016lrs} combined classical Yang-Mills evolution  with hadronization using the Lund string fragmentation model. A  mass ordering similar to hydrodynamics was seen for  both $\left<\ptt\right>$ and $v_2$. This indicates that the hadron specied dependence is a more generic effect of hadronization, and should thus not by itself be taken as a confirmation of hydrodynamic behavior.

In any case, for small systems azimuthal multiparticle correlations can be generated by both hydrodynamical flow, and by coherent effects in the color fields of the colliding projectiles. The first mechanism becomes parametrically more important as the system grows larger and its lifetime increases. The second one, on the other hand, is relativley larger for smaller systems, where the fluctuations in the color fields are not averaged out as efficiently~\cite{Gelis:2009wh}. Quantitatively sorting out the relative importance of these two effects will still require more careful theoretical work, and a simultaneous comparison to many experimental observables.

\section{Hard and electromagnetic probes}

Perhaps the largest consequence for heavy ion physics of the increase in center of mass energy from  RHIC to the LHC is the presence of very high energy jets. The combinatorial background in a heavy ion collision is much higher that in proton-proton events, so a sufficiently high $\ptt$ is required to meaningfuly cluster produced particles into jets. Jets, and more generally other hard probes of the collision, do not form a part of the plasma. In stead the jet should be thought of as an external calibrated probe whose production cross section can be calculated from first principles and which interacts with the QCD matter medium as it passes through it. 

The LHC measurements of jet cross sections, dijet correlations etc. have prompted a flurry of theoretical advances that are impossible to describe here in any very significant detail. The basic qualitative picture emerging from these studies (see e.g.~\cite{MehtarTani:2010ma,MehtarTani:2011tz,Blaizot:2012fh,Blaizot:2013vha,Blaizot:2013hx,CasalderreySolana:2012ef,Kurkela:2014tla}) is that of the leading particle losing energy due to scatterings from the medium. Because of a loss of color coherence caused by interactions with the medium, this cascade can be described in terms of a probabilistic branching process. Only when the energy has cascaded into very soft particles at the medium scale, does the energy scatter outside of the jet cone into very large angles. The signature of this process, indeed seen in the experiments,  is that the jet that loses energy to very large angle scattering outside the jet cone, but  the fragmentation of the part of the jet remaining inside the cone is relatively  unmodified by the medium. 

\begin{wrapfigure}{r}{0.5\textwidth}
\includegraphics[width=0.5\textwidth]{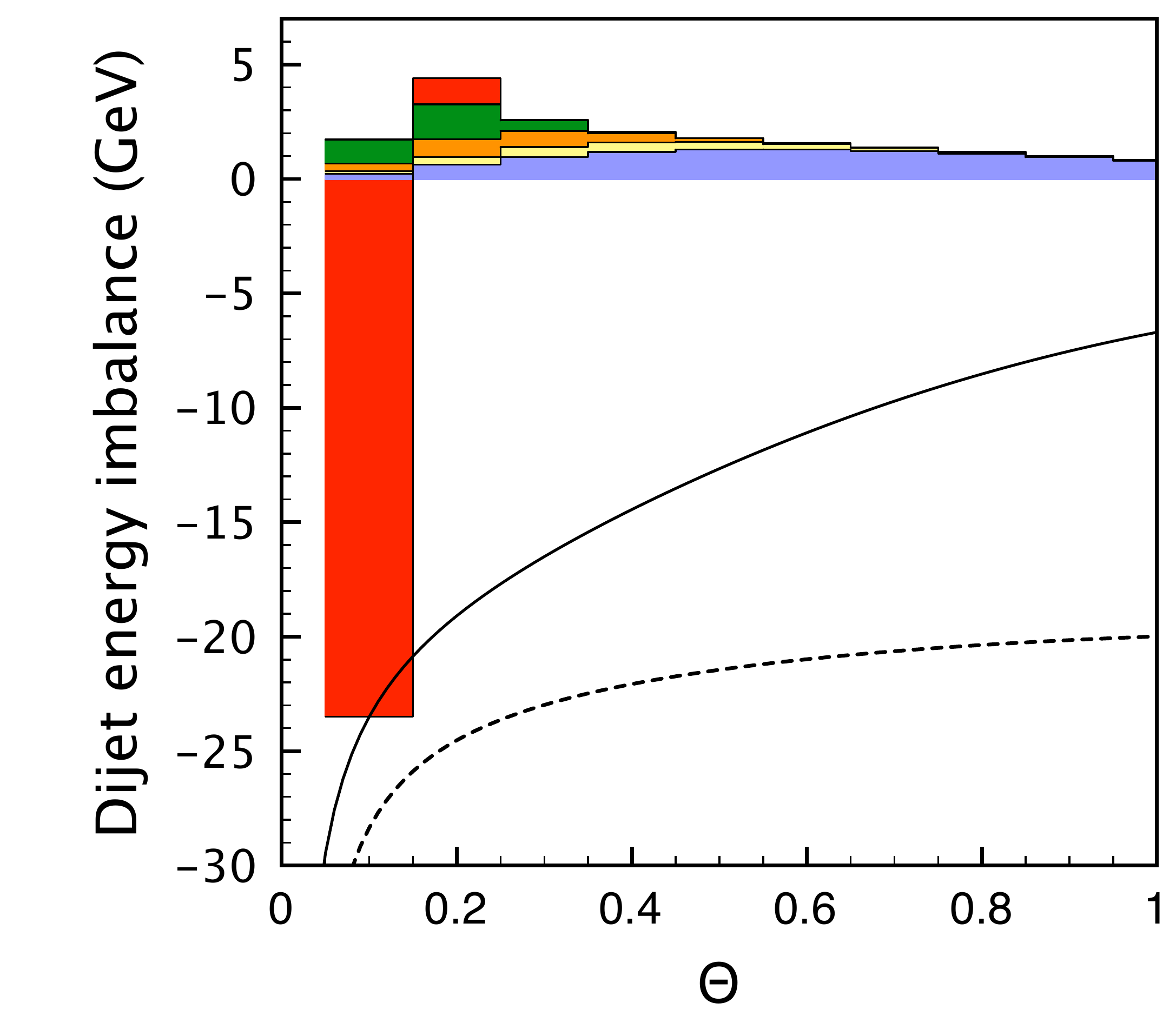}
\caption{Modification energy within a jet in a nucleus-nucleus collision compared to proton-proton collisions at different angles from the cone axis: energy is lost by radiation by the particles colinear to the jet axis and  transferred by large angle scattering to large angles, from~\cite{Blaizot:2014ula}.
} 
\end{wrapfigure}
The medium itself is parametrized by a ``transport coefficient'' $\hat{q}$, which parametrizes the transverse momentum squared accumulated per unit length by a traversing fast particle. Recently first NLO corrections to the jet-medium interaction have been calculated, leading to a renormalization of this coefficient that absorbs the leading logarithmic dependence on the medium length of these corrections~\cite{Wu:2011kc,Liou:2013qya,Blaizot:2013vha}.

As has been emphasized several times in this conference~\cite{Sjostrand:2016bif}, these theory developments do not become fully meaningful until they are impletented in Motne Carlo event generators. There are now several competing groups working on porting these new theoretical developments into generators, either new or constructed as modifications of existing vacuum jet Monte Carlos~\cite{Renk:2008pp,Armesto:2009fj,Schenke:2009gb,Zapp:2013vla,Bravina:2016bei}.

\begin{figure}[t]
\resizebox{\textwidth}{!}{
\includegraphics[height=4cm]{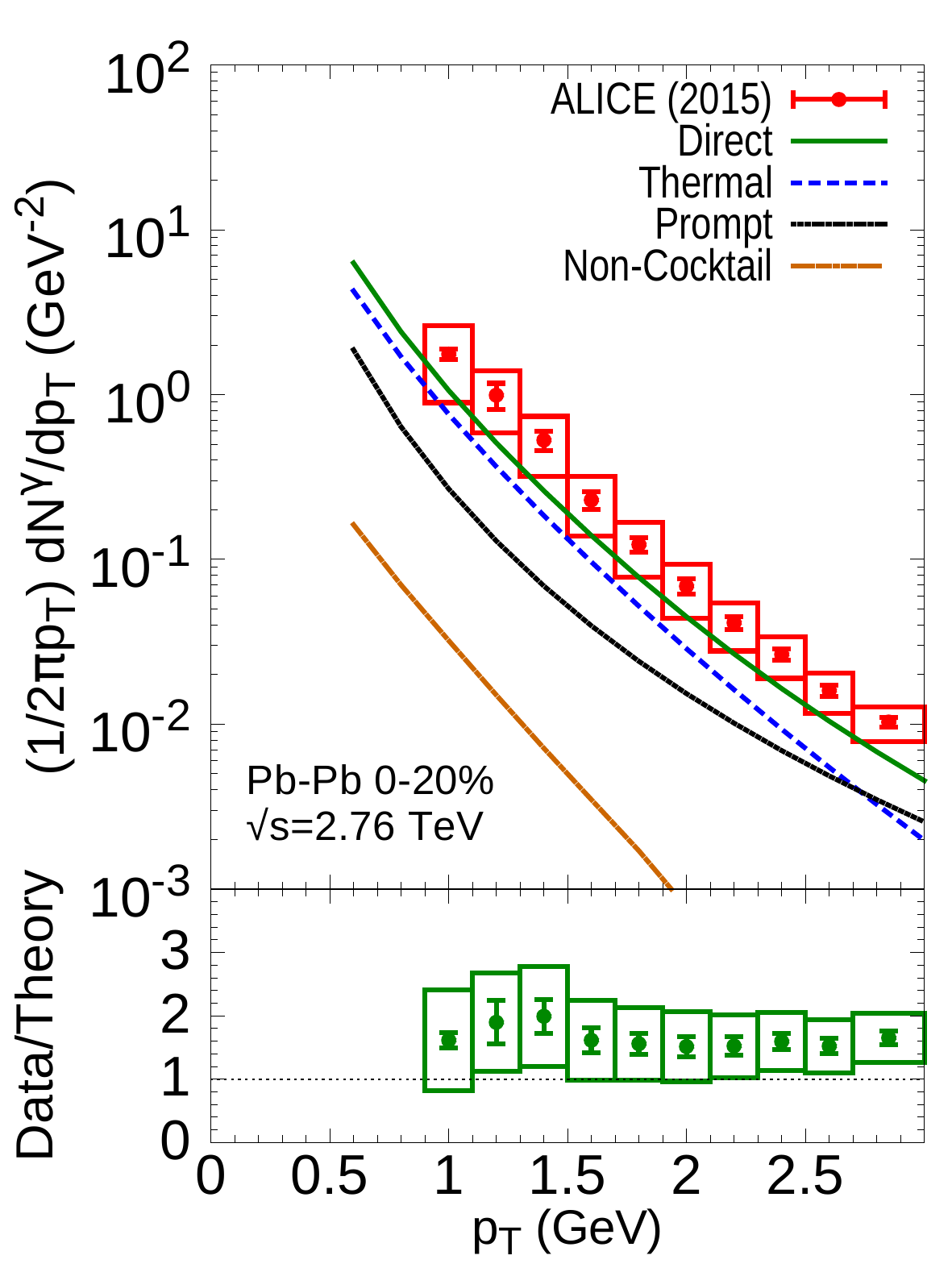}
\includegraphics[height=4cm]{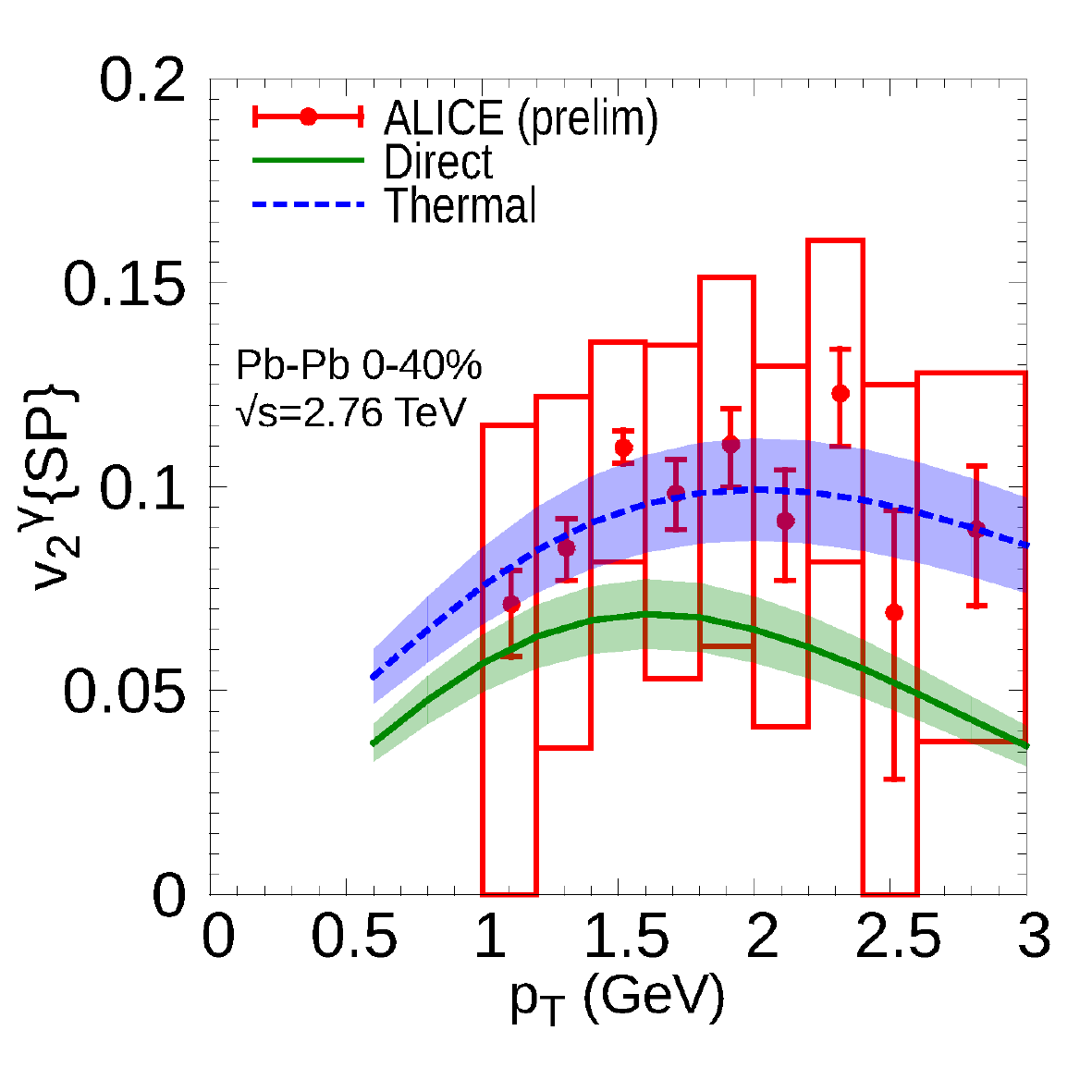}
}
\caption{Comparison of photon yields and elliptic flow coefficients with ALICE data, from the calculation~\cite{Paquet:2015lta}.} 
\label{fig:phot}
\end{figure}

Photons and dileptons are a different type of a perturbative probe of the deconfined medium. Unlike jets, they escape from the fireball practically without interactions. In stead, they can be produced directly by the medium as thermal radiation in addition to the production mechanisms that are present also in proton-proton collisions. Thermal photons are important as the only \emph{direct} probe emitted from the deconfined plasma stage. On the other hand, the interpretation of the photon measurements is complicated by the presence of several other photon sources, in particular ``direct photons'' and hadronic decays. Here one has very recently been edging towards a solution of a major outstanding issue for several years referred to as the ``photon $v_2$ puzzle''.  In short, calculations have found it very difficult to simultaneously describe the observed large photon yield and the rather large photon elliptic flow $v_2$. The large yield by itself would naturally result from a higher thermal photon emission rate from the early plasma phase. But when the plasma is hot, it is not yet flowing very fast, and these early time photons have a very small $v_2$. The large observed $v_2$, on the other hand, seems to point towards a dominance of photon emission from the late hadronic stage when the matter has a large azimuthal asymmetry, but then it becomes difficult to explain the yield. The solution to this issue seems not to reside in a  single improvement, but depends on various factors such as the hadronic thermal photon emission rates and the effects of bulk and shear viscosity on the hydrodynamical evolution. The results of a recent calculation~\cite{Paquet:2015lta} including both emissions from the hydrodynamical medium and perturbative ``direct'' processes is shown in Fig.~\ref{fig:phot}.

\begin{wrapfigure}{r}{0.5\textwidth}
 \includegraphics[width=0.5\textwidth]{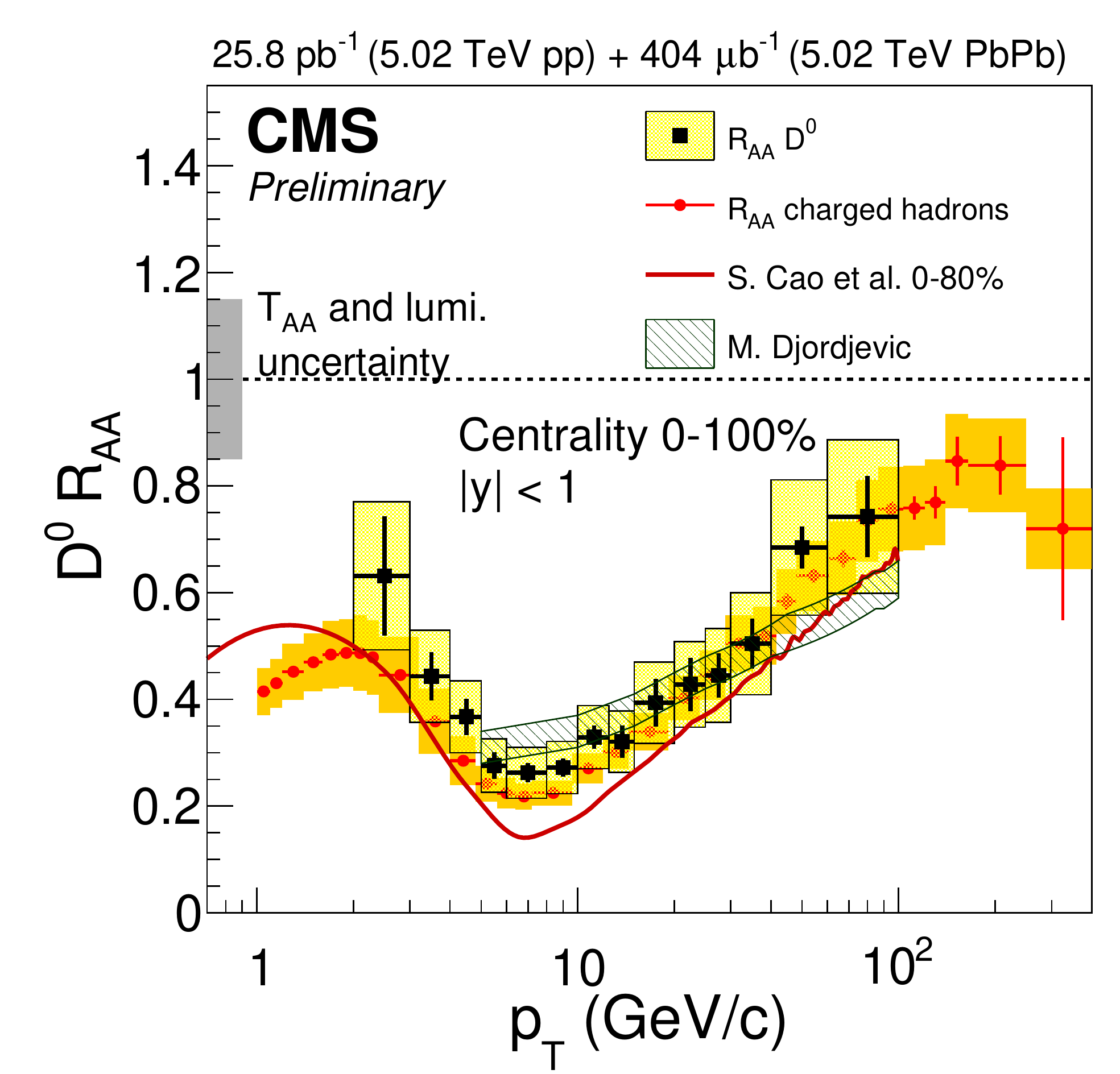}
\caption{CMS result~\cite{CMS:2016nrh} for the suppression of $D$-meson production with respect to the proton-proton baseline scaled by  nuclear geometry, i.e. $R_{AA}$.}
\label{fig:draa}
\end{wrapfigure}

The term ``open heavy flavor'' refers to heavy quarks not bound in $c\bar{c}$ or $b\bar{b}$ bound states. To a first approximation a heavy quark meson such as the $D$ is a proxy for a single heavy quark which is created in a perturbative process and then has to propagate in the plasma, before hadronizing. The LHC experimental result is that the nuclear modification of the spectrum of heavy quark mesons at semihard and high transverse momenta is very similar to that of light hadrons, as demonstrated e.g. in Fig~\ref{fig:draa}. While at first sight this might not seem so surprising, it is made nontrivial by the fact that the  theoretical treatments of  heavy and light mesons invoke very different physics. For light hadrons the spectrum is usually thought of in terms of a two component picture of an exponentially falling thermal one and a power law perturbative tail from the fragmentation of quenched jets. This results in the characteristic first falling, then increasing nuclear modification ratio $R_{AA}$ as a function of $\ptt$. Heavy quarks, on the other hand, are not obviously produced thermally, but do interact with the medium, being ``pushed around'' by thermal motion and following the flow of the plasma. In spite of this very different physical picture, some recent calculations are able to reproduce rather well the features seen in the data.

\begin{figure}[t]
\centerline{\includegraphics[width=0.7\textwidth]{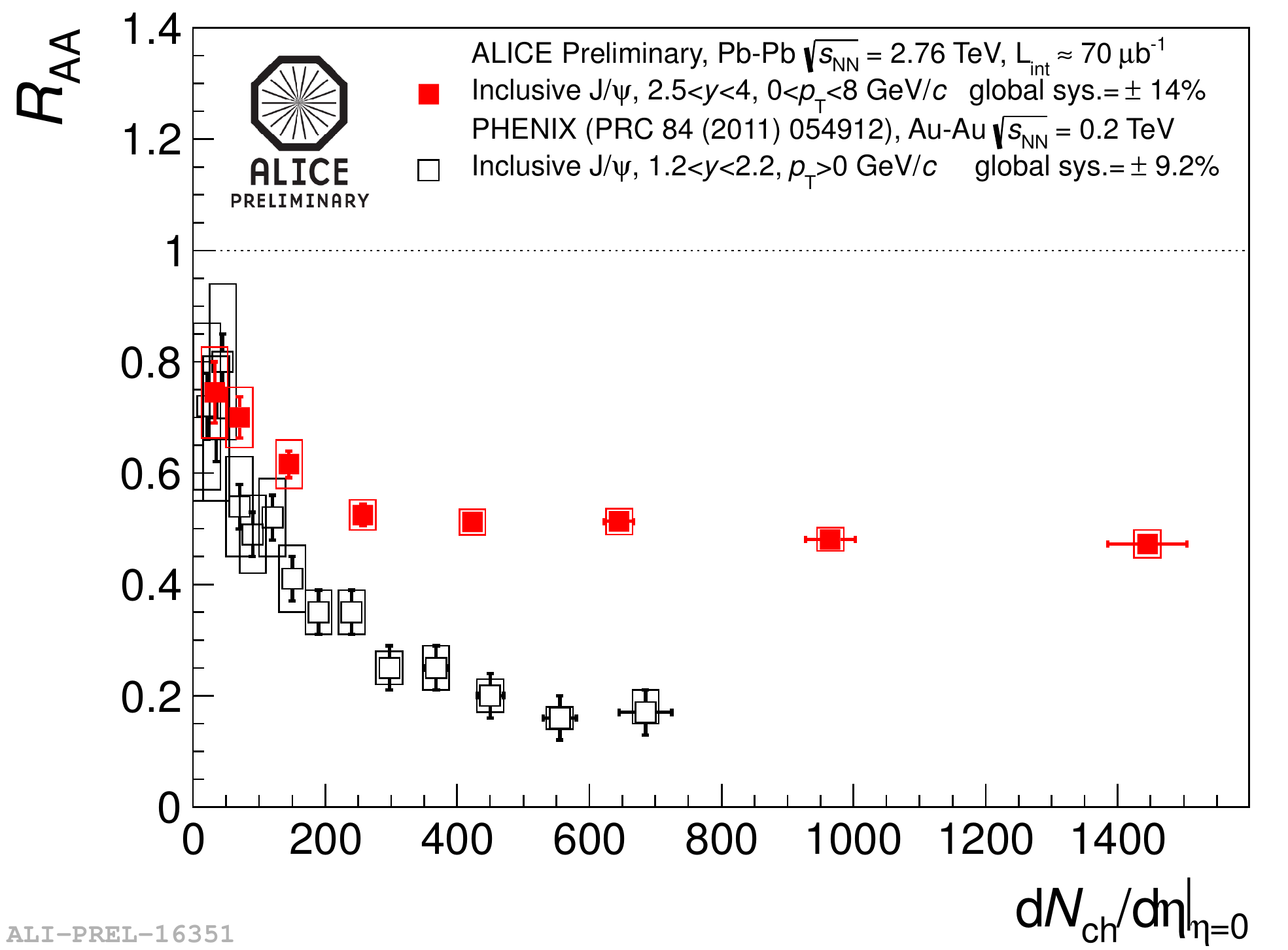}}
\caption{ALICE and PHENIX results for $J/\psi$ nuclear suppression at LHC and RHIC energy~\cite{Maire:2013ad}.}
\label{fig:jpsiraa}
\end{figure}

Quarkonia, e.g. the ever-important $J/\psi$, are the quintessential semihard QCD probe. They are perturbative enough to be at least qualitatively understood in weak coupling but nevertheless light enough to be copiously produced and strongly interacting. Quarkonia have long been advertised as the QGP thermometer~\cite{Matsui:1986dk}, which would sequantially melt and disappear from the measurements as the temperature of the plasma is raised with higher collision energy. In reality it turns out that the thermometer is more complicated than thought: as shown in Fig~\ref{fig:jpsiraa} $J/\psi$ production is suppressed less, not more at LHC energies compared to RHIC. In order to fully understand the systematics of the QGP thermometer one needs to understand very accurately quarkonium production in proton-proton collisions, the nuclear effects on the production, the bound state dissociation in a thermal medium using the lattice, the propagation in the medium that leads to a boost with the flow and affects the $\ptt$ spectrum, and finally contributions from a regeneration of new  $c\bar{c}$ bound states from thermal charm quark pairs in the medium. The community is making progress in all of these aspects, but there is still much work to do befor the ``simple'' probe of the QGP is understood.

\section{Conclusions}

To conclude, there has been a huge progress in systematically extracting fundamental properties of quark-gluon matter from experiment. The main theoretical tool in this work is a global description of soft observables using relativistic hydrodynamical simulations. These allow for a simultaneous computation of quantities that can directly be compared to experimental results. 

Nevertheless, there is still much work to do to beter understand QCD in a dense and hot environment. In nucleus-nucleus collisions jets and heavy quarks can be used to understand the properties of the medium by their interactions with energetic colored particles, and electromagnetic probes provide additional independent constraints on the spacetime evolution of the plasma.

Proton-nucleus and proton-proton collisions have turned out to be much more interesting than was believed before the LHC started. They have to be studied carefully to understand ``cold'' nuclear matter effects for the production of perturbative probes: high $\ptt$ particles, heavy quarks and electromagnetic probes. In addition to serving as a control experiment, these small collision systems have also revealed strong effects pointing towards collective behavior. Understanding the precise nature of this collectivity is and will be an active and interesting topic for research in the years to come.
\paragraph{Acknowledgements}
 This work has been supported by the Academy of Finland, projects 
267321 and 273464. 

\bibliography{spires}
\bibliographystyle{JHEP-2modlong}

\end{document}